\documentclass{article}
\usepackage{amsmath}

\input{tcilatex}
\begin{document}

\begin{center}
{\large \ The geodesics structure of Schwarzschild electromagnetic black hole%
}{\LARGE \ }

A. Al-Badawi, M.Q. Owaidat and S. Tarawneh

Department of Physics, Al-Hussein Bin Talal University, P. O. Box: 20,
71111, Ma'an, Jordan

\bigskip Department of Mathmatics, Al-Hussein Bin Talal University, P. O.
Box: 20, 71111, Ma'an, Jordan

Email: ahmadalbadawi@hotmail.com , Owaidat@ahu.edu.jo and
math\_10039@yahoo.com

{\Large Abstract}
\end{center}

The geodesic equations are considered in static mass imbedded in a uniform
electromagnetic field. Due to electromagnetic field horizon shrinks and
geodesics are modified. By analyzing the behavior of the effective
potentials for the massless and massive particle we study the radial and
circular trajectories. Radial geodesics for both photons and particles are
solved exactly. It is shown that a particle falls toward the horizon in a
finite proper time slows down so that the particle reaches the singularity
in longer time than Schwarzschild case. Timelike and null circular geodesics
are investigated. We have shown that, there is no stable circular orbits for
photons, however stable and unstable second kind orbits are exists for
massive particle. An exact analytical solution for the innermost circular
orbits $\left( ISCO\right) $\ has been obtained. It has been shown that the
radius of $ISCO$\ shrinks due to the presence of electromagnetic field.

\section{Introduction}

A solution that unify the two well known solutions of General Relativity
(GR), Schwarzschild (S) and Bertotti-Robinson (BR) solution [1-2], in a
common metric has been obtained previously [3]. This metric represents a S
mass embedded in an external em BR universe. We shall \ phrased hereafter
the solution as the Schwarzschild Electromagnetic Black Hole (SEBH). The
SEBH solution is obtained by solving the Einstein-Maxwell (EM) field
equations with S and BR solutions as boundary conditions. The resulting
solution is partly transformable to the Reissner-Nrdstr\"{o}m (RN) solution
by Birkhoff theorem. From mathematical point of view, SEBH is nothing but
the interpolation of two exact solutions of EM field equations. Recall that,
considering an isolated system in a uniform external em field has been
studied by many researches. To be more precise, we are concerned with the
SBH with an em field [4-7].

In Newtonian gravity of flat space a chargless particle do not interact with
em field. However, through the curvature of the space in GR a chargless mass
do interacts with the surrounding em field. Therefore as a result of
embedding S in an external em BR universe, horizon and particle geodesics
are modified. The main idea of this paper is to investigate both the radial
and circular geodesics in SEBH. In particular, we want to expose the effect
of coupling S with BR on the trajectories of the null and timelike geodesics.

\ On the other hand, the geodesic structure has been studied earlier for SBH
and RNBH [8], Kerr-Taub-NUT BH [9] and recently Kerr Newman-Taub-NUT [10].
The geodesics structure in S anti-deSitter BH was studied in [11-13]. The
authors in Ref. [13] have found all possible motions which are allowed by
the energy levels, radial and circular trajectories were exactly evaluated
for both geodesics. Geodesic structure of test particle in Bardeen spacetime
was studied in [14]. The study of test particle in RN geometry was
investigated in [15]. Analysis of the effective potential in RN de Sitter \
and Kerr de Sitter spacetime were investigated in [16]. All possible
geodesic motions in the extrema S de Sitter geometry was realized in [17].
Analytical solutions and applications of the geodesic equation in S (anti)
de Sitter spacetimes was realized in [18]. Recently, the effect of the
Rindler parameter on the trajectories of the timelike and null geodesics has
been investigated [19] and compared with S geodesics. It is shown that the
Rindler acceleration tends to confine the radial geodesics. Motion of
massive particles around a rotating BH in a braneworld has been studied in
[20]. Where in [20] they have obtain an exact analytical solution for the
dependence of the radius of the innermost stable circular orbits (ISCO).
Recall that the ISCO represents the minimum possible radius of stable
circular orbit. It is known that for SBH the $r_{ISCO}$ is 6M [21-22] and
for Kerr spacetime is 9M. The ISCO was analysised in Kerr geometry for a
spinless particle [23-24]. Whereas the ISCO for the motion of a spin test
particles in S and Kerr geometries was investigated in [25].

In this work we address the study of null and timelike geodesic in the
background of SEBH. We analysis the effective potentials and found all
possible orbits allowed. Radial geodesics for both photon and massive
particles were solved exactly. It is shown that, particles falling toward
the horizon in a finite proper time in SEBH take longer than SBH case.
Timelike and null circular geodesics are studied.\ We have shown that there
are no stable circular orbits for photons and all possible circular orbits
obtained are similar to the circular orbits in RNBH. Finally, we have
obtained an exact analytical solution for the ISCO in SEBH.

The motivation of this paper is to expose the modified geodesics of SEBH\
solution which is a result of S immersed in an em universe. The timelike and
null trajectories are very important to understand the various physical
phenomenon i.e. perihelion shifts, gravitational redshifts, light bending
and different observables like Lense-Thirrring effect, gravitational time
delay etc. All are phenomenons related to the geodesics. Besides the
physical motivation reasons it is also of mathematical important to derive
an explicit analytical solution of the geodesics equations in the back
ground of SEBH. To achieve our aim we will follow the procedure used in S
metric [8], and other studies [26-31]. Our paper is organized as follows: in
Sect. 2, we review of SEBH metric and derive the geodesic equations that
give us a complete description of the motion of the particle with SEBH
metric. In Sect. 3, the radial geodesic for null and timelike particle are
solved exactly. In Sect. 4, the effective potential is analyzed and all
possible motions of circular orbits for null and massive particle are
determined. Also in this Sect. we have obtained an exact analytical solution
for ISCO. Finally we make concluding remarks.

\section{Structure of SEBH\ solution and derivation of geodesic equations}

We begin by recalling the SEBH metric which represents the non-linear
superposition of the S solution and the BR solution [5,32]:

\begin{equation}
ds^{2}=-f\left( r\right) dt^{2}+\frac{1}{f\left( r\right) }%
dr^{2}+r^{2}\left( d\theta ^{2}+\sin ^{2}\theta d\phi ^{2}\right) ,
\end{equation}

where the lapes function $f\left( r\right) $, has the following form%
\begin{equation}
f\left( r\right) =1-\frac{2M}{r}+\frac{M^{2}}{r^{2}}\left( 1-a^{2}\right) .
\end{equation}%
in which, $M$ is the \ S mass coupled to\ an external em field and $a\left(
0<a\leq 1\right) $ is the external parameter. When $a=1,$ the metric reduces
to S solution. The case $\left( a=0\right) ,$ which is excluded, is the
extremal RN which is transformable to BR metric. As a result of the coupling
of S and BR, the horizon shrinks to a significant degree. The horizon is
given by%
\begin{equation}
r_{h}=M\left( 1+a\right) .
\end{equation}%
It is clear that $r_{h}\leq 2M$, since $\left( 0<a\leq 1\right) .$ By using
the Newman-Penrose (NP) formalism [33], we choose the null tetrad 1-form $%
\left( l,n,m,\overline{m}\right) $ in terms of the metric functions that
satisfies the orthogonality conditions, $\left( l.n=-m\text{.}\overline{m}%
=1\right) $. Where a bar over a quantity denotes complex conjugation. We
assume the null tetrad frame to be given by \bigskip 
\begin{equation*}
l_{\mu }=dt-\frac{dr}{f\left( r\right) },
\end{equation*}%
\begin{equation*}
n_{\mu }=\frac{1}{2}f\left( r\right) dt+\frac{1}{2}dr,
\end{equation*}%
\ 
\begin{equation}
m_{\mu }=\frac{-r}{\sqrt{2}}(d\theta +i\sin \theta d\phi ),
\end{equation}%
and 
\begin{equation*}
l^{\mu }=f\left( r\right) dt+dr,
\end{equation*}%
\begin{equation*}
n^{\mu }=\frac{1}{2}dt-\frac{1}{2}f\left( r\right) dr,
\end{equation*}%
\ 
\begin{equation}
m^{\mu }=\frac{1}{\sqrt{2}r}(d\theta +\frac{i}{\sin \theta }d\phi ),
\end{equation}

We determine the nonzero NP complex spin coefficients as, 
\begin{equation}
\rho =-\frac{1}{r},\qquad \mu =\frac{-f\left( r\right) }{2r},\qquad \gamma =%
\frac{1}{4}\frac{df\left( r\right) }{dr},\qquad \alpha =-\beta =\frac{-\cot
\theta }{2\sqrt{2}r}.
\end{equation}%
We also obtain the Weyl and Ricci scalars  
\begin{equation*}
\Psi _{2}=\frac{-Ma}{R^{6}}\left( a^{2}r^{3}+\left( 1-a\right) \left[
Ma\left( 2-a\right) r^{2}+M^{2}\left( 1-a\right) \left( 1-2a\right)
r-M^{3}\left( 1-a\right) ^{2}\right] \right) 
\end{equation*}%
\begin{equation}
\Phi _{11}=\frac{M^{2}\left( 1-a^{2}\right) }{2R^{4}},
\end{equation}%
\ where 
\begin{equation}
R=ar+\left( 1-a\right) M.
\end{equation}

This metric is Petrov type-D spacetime. It is seen that metric (1) satisfies
all the required limits as boundary conditions:

\begin{eqnarray}
\ BR\ solution &\rightarrow &\left( a=0,M=1\right) \rightarrow \Psi
_{2}=0,\Phi _{11}=\frac{1}{2}, \\
\ S\,solution &\rightarrow &\left( a=1,R=r\right) \rightarrow \Psi _{2}=%
\frac{-M}{r^{3}},\Phi _{11}=0,  \notag \\
SEBH\ solution &\rightarrow &\left( 0<a\leq 1\right) \rightarrow \Psi
_{2}\neq 0\neq \Phi _{11}.  \notag
\end{eqnarray}

\bigskip Now we will obtain the geodesics equations which can be derived
from the Lagrangian 
\begin{equation}
\mathcal{L}=\frac{1}{2}g_{\mu \nu }\frac{dx^{\rho }}{ds}\frac{dx^{\sigma }}{%
ds},
\end{equation}%
where $s$ is some affine parameter along the geodesics. For SEBH\ the
Lagrangian is 
\begin{equation}
2\mathcal{L}=-f\left( r\right) \dot{t}^{2}+\frac{1}{f\left( r\right) }\dot{r}%
^{2}+r^{2}\left( \dot{\theta}^{2}+\sin ^{2}\theta \dot{\phi}^{2}\right) ,
\end{equation}%
in which a dot denotes a differentiation with respect to $s.$ Since the
Lagrangian is independent of $\left( t,\phi \right) $ the corresponding
conjugate momenta are conserved, therefore%
\begin{equation}
E=\frac{d\mathcal{L}}{d\overset{\cdot }{t}}=g_{tt}\frac{dt}{ds}=-f\left(
r\right) \frac{dt}{ds}=\text{constant}
\end{equation}%
and 
\begin{equation}
\ell =\frac{d\mathcal{L}}{d\overset{\cdot }{\phi }}=r^{2}\sin ^{2}\theta 
\overset{\cdot }{\phi }=\text{constant.}
\end{equation}%
Let us choose $\theta =\pi /2$ and $\dot{\theta}=0.$ This means that the
motion is restrict to the equatorial plane. Since we have two constants of
motion Eqs. (12) and (13) the geodesic equation reduces to 
\begin{equation}
\left( \frac{dr}{d\phi }\right) ^{2}=\frac{r^{4}}{\ell ^{2}}\left(
E^{2}-f\left( r\right) \right) \left( \epsilon +\frac{\ell ^{2}}{r^{2}}%
\right) .
\end{equation}%
We can obtain the $r$ equation as a function of $s$ and $t$%
\begin{equation}
\left( \frac{dr}{ds}\right) ^{2}=E^{2}-f\left( r\right) \left( \epsilon +%
\frac{\ell ^{2}}{r^{2}}\right) ,
\end{equation}%
\begin{equation}
\left( \frac{dr}{dt}\right) ^{2}=\frac{1}{E^{2}}\left( E^{2}-f\left(
r\right) \left( \epsilon +\frac{\ell ^{2}}{r^{2}}\right) \right) \left(
f\left( r\right) \right) ^{2}.
\end{equation}%
The above three equations (14-16) give us a complete description of the
motion of the particle. let us note that, for timelike we choose $\epsilon
=1 $, and for null geodesics we choose $\epsilon =0.$ Eq.(15) can be cast
into a familiar form of equation of motion for a unit mass test particle as 
\begin{equation}
\frac{1}{2}\left( \frac{dr}{ds}\right) ^{2}+V_{eff}\left( r\right)
=\varepsilon _{eff},
\end{equation}%
with an effective energy $\varepsilon _{eff}=\frac{1}{2}E^{2}$ and effective
potential 
\begin{equation}
V_{eff}\left( r\right) =\frac{1}{2}\left( 1-\frac{2M}{r}+\frac{M^{2}}{r^{2}}%
\left( 1-a^{2}\right) \right) \left( \epsilon +\frac{\ell ^{2}}{r^{2}}%
\right) .
\end{equation}%
For a real classical region, $r$ is limited by the constraint \bigskip\ $%
\varepsilon _{eff}\geq V_{eff}\left( r\right) .$ As usual, we introduce $%
u=1/r$ and obtain%
\begin{equation}
\left( \frac{du}{d\phi }\right) ^{2}=\frac{E^{2}}{\ell ^{2}}-\left(
1-2Mu+M^{2}\left( 1-a^{2}\right) u^{2}\right) \left( \frac{\epsilon }{\ell
^{2}}+u^{2}\right) =g\left( u\right) .
\end{equation}

\bigskip

\section{Radial Geodesics}

In this section we shall study the motion of the radial geodesics in zero
angular momentum (i.e. $\ell =0$) which means that the motion will remain in
the $\phi =$ constant plane and particle will move radially.

\subsection{Null geodesics $\protect\epsilon =0.$}

In the case of a massless particle (photon) Eq. (15) simplifies to%
\begin{equation}
\frac{dr}{ds}=\pm E,
\end{equation}%
where (-) stands for ingoing photon and (+) stands for outoing photon. As in
S case the solution of Eq. (20) is $r=\pm E\mathcal{\tau }+r_{0},$ where $%
r_{0}$\ corresponds to the initial position of the photon. In order to
obtain the equation of motion in terms of $t,$ we use Eq.(12)\bigskip\ and
after some modification we obtain 
\begin{equation}
\frac{dr}{dt}=\pm \left( 1-\frac{2M}{r}+\frac{M^{2}}{r^{2}}\left(
1-a^{2}\right) \right) .
\end{equation}%
The solution is given by%
\begin{eqnarray}
\pm \left( t-t_{0}\right) &=&r+M\log \left( \frac{\left( r-r_{h}\right)
\left( r-M\left( 1-a\right) \right) }{M^{2}\left( 1-a^{2}\right)
-2Mr_{0}+r_{0}^{2}}\right)  \notag \\
&&+\frac{M\left( 1+a^{2}\right) }{a}\left[ arc\tan \left( \frac{M-r}{aM}%
\right) -arc\tan \left( \frac{M-r_{0}}{aM}\right) \right] .
\end{eqnarray}%
Where $r_{0}$ and $t_{0}$ are the initial position and time of the massless
particle while $t$ is the time measured by the distance observer.\bigskip\
It is seen that, when $r\rightarrow r_{h}$, then $t\rightarrow \infty $ as
in the case of SBH.

\subsection{Timelike geodesics $\protect\epsilon =1.$}

Now we will study the equations of motion of a massive particle, therefore
Eq. (15) becomes%
\begin{equation}
\left( \frac{dr}{ds}\right) ^{2}=E^{2}-1+\frac{2M}{r}-\frac{M^{2}}{r^{2}}%
\left( 1-a^{2}\right) .
\end{equation}%
Choosing $s$ to be the proper time $\mathcal{\tau }$, then Eq. (23) can be
rewritten as 
\begin{equation}
\frac{d^{2}r}{d\mathcal{\tau }^{2}}=\frac{M^{2}}{r^{3}}\left( 1-a^{2}\right)
-\frac{M}{r^{2}}.
\end{equation}%
In radial motion the effective potential is given by 
\begin{equation}
V_{eff}\left( r\right) =\frac{1}{2}\left( 1-\frac{2M}{r}+\frac{M^{2}}{r^{2}}%
\left( 1-a^{2}\right) \right) .
\end{equation}%
The evolution of the effective potential for the radial particles at
different values of the external parameter $a$ is depicted in Fig. 1. We
observe from Fig. 1 that for increasing the external parameter $a$ the
radial orbits become more stable therefore all radial orbits in SEBH are
stable. This is because the effective potential Eq. (25) always has a
minimum for any value of $a$ except when $a=1$ which corresponds to S
spacetime. We also note that Fig. 1 shows there is no upper bound for the
radial motion of the particle  which implies that particle in SEBH can
escape to infinity as in the case of SBH.

Now let us consider the particle falling towards the center from a finite
distance $r_{0}$ with zero initial velocity where the starting distance is
related to the constant $E$ by $E^{2}=1-\frac{2M}{r_{0}}+\frac{M^{2}}{%
r_{0}^{2}}\left( 1-a^{2}\right) $. Making the change of variable $%
r=r_{0}\cos ^{2}\eta /2$, we obtain 
\begin{equation}
\left( \frac{dr}{d\mathcal{\tau }}\right) ^{2}=M^{2}\left( 1-a^{2}\right)
\left( \frac{1}{r_{0}^{2}}-\frac{1}{r^{2}}\right) +2M\left( \frac{1}{r}-%
\frac{1}{r_{0}}\right) .
\end{equation}%
Therefore in terms of $\eta $ the equation to be integrated together with $%
\frac{dr}{d\eta }=-r_{0}\sin \frac{\eta }{2}\cos \frac{\eta }{2}$ is 
\begin{equation}
\left( \frac{dr}{d\mathcal{\tau }}\right) ^{2}=\tan ^{2}\frac{\eta }{2}%
\left( 1-E^{2}-\frac{M^{2}\left( 1-a^{2}\right) }{r_{0}^{2}\cos ^{2}\frac{%
\eta }{2}}\right) .
\end{equation}%
For infalling \bigskip particles we consider%
\begin{equation}
\frac{dr}{d\mathcal{\tau }}=-\tan \frac{\eta }{2}\sqrt{\frac{2M}{r_{0}}-%
\frac{M^{2}}{r_{0}^{2}}\left( 1-a^{2}\right) -\frac{M^{2}\left(
1-a^{2}\right) }{r_{0}^{2}\cos ^{2}\frac{\eta }{2}}}
\end{equation}
Eq. (28) can be rewritten as 
\begin{equation}
\frac{d\mathcal{\tau }}{d\eta }=\frac{r_{0}\cos ^{2}\frac{\eta }{2}}{\sqrt{%
\frac{2M}{r_{0}}-\frac{M^{2}}{r_{0}^{2}}\left( 1-a^{2}\right) -\frac{%
M^{2}\left( 1-a^{2}\right) }{r_{0}^{2}\cos ^{2}\frac{\eta }{2}}}}.
\end{equation}%
On integration, in terms of the variable $r\left( \eta _{0}=0\text{ if }%
r=r_{0}\text{ and }\mathcal{\tau }_{0}=0\right) $ this equation gives%
\begin{equation}
\mathcal{\tau }=\frac{2M}{R_{0}^{3/2}}Arc\tan \left[ \frac{\sqrt{2R_{0}}\sin
\left( \eta /2\right) }{\sqrt{R_{0}\left( 1+\cos \left( \eta \right) \right) 
}}\right] +\frac{\sqrt{R_{0}\left( 1+\cos \left( \eta \right) \right) }}{%
\sqrt{2}R_{0}}r_{0}\sin \left( \eta /2\right) ,
\end{equation}%
\newline
where 
\begin{equation}
R_{0}=\frac{2M}{r_{0}}-\frac{M^{2}}{r_{0}^{2}}\left( 1-a^{2}\right) 
\end{equation}%
Note that, when we substitute the limiting case $a=1,$ the exact analytical
expression Eq. (30) reduces to%
\begin{equation}
\mathcal{\tau }=\left( \frac{r_{0}^{3}}{8M}\right) ^{1/2}\left( \eta +\sin
\eta \right) ,
\end{equation}%
which coincides exactly with the S spacetime. Eq. (30) is plotted in Fig.
(3) together with SBH case (see Fig. 2) for comparison. Coparing the two
figures we see that particle in SEBH takes longer time than in the case of
SBH. Therefore as a result of coupling the S with BR, a particle falling
towards the horizon disaccelerates so that the particle takes longer time to
reach the singularity.

The situation is different when we consider the equation of the trajectory
in the coordinate time $t.$ Using Eqs. (27) and (29) we obtain%
\begin{equation}
dt=\dint \frac{Er_{0}^{3}\cos ^{7}\frac{\eta }{2}d\eta }{\left( r_{0}\cos
^{2}\frac{\eta }{2}-r_{h}\right) \left( r_{0}\cos ^{2}\frac{\eta }{2}%
-m\left( 1-a\right) \right) \sqrt{R_{0}\cos ^{2}\frac{\eta }{2}-\frac{M^{2}}{%
r_{0}^{2}}\left( 1-a^{2}\right) }}.
\end{equation}%
In the limit, as $\eta \rightarrow \eta _{h},$ $t\rightarrow \infty $ which
is consistent with the S case. We notice that the conserved energy of the
particle can be found from Eq. (16), namely%
\begin{equation}
0=\frac{1}{E^{2}}\left( E^{2}-1+\frac{2M}{r_{0}}-\frac{M^{2}}{r_{0}^{2}}%
\left( 1-a^{2}\right) \right) \left( 1-\frac{2M}{r_{0}}+\frac{M^{2}}{%
r_{0}^{2}}\left( 1-a^{2}\right) \right) ^{2}
\end{equation}%
which implies that 
\begin{equation}
E^{2}=1-\frac{2M}{r_{0}}+\frac{M^{2}}{r_{0}^{2}}\left( 1-a^{2}\right) .
\end{equation}

\bigskip

\section{Circular geodesics}

We shall study in this section the circular motion $\left( \ell \neq
0\right) $ of mass and massless particle. The conditions of the occurrence
of circular orbits are $g\left( u\right) =0,$ and $g^{\prime }\left(
u\right) =0.$ Applying these conditions we obtain%
\begin{equation}
\frac{E^{2}}{\ell ^{2}}-\left( 1-2Mu+M^{2}\left( 1-a^{2}\right) u^{2}\right)
\left( \frac{\epsilon }{\ell ^{2}}+u^{2}\right) =0,
\end{equation}%
and 
\begin{equation}
\frac{d}{du}\left( \frac{E^{2}}{\ell ^{2}}-\left( 1-2Mu+M^{2}\left(
1-a^{2}\right) u^{2}\right) \left( \frac{\epsilon }{\ell ^{2}}+u^{2}\right)
\right) =0.
\end{equation}%
From the above Eqs. (36) and (37) we can obtain the expressions for the
energy $E$ and the angular momentum $\ell $ of the particle 
\begin{equation}
E^{2}=\frac{\epsilon \left( 1-2Mu_{c}+M^{2}u_{c}^{2}\left( 1-a^{2}\right)
\right) ^{2}}{1-3Mu_{c}+2M^{2}u_{c}^{2}\left( 1-a^{2}\right) },
\end{equation}%
\begin{equation}
\ell ^{2}=\frac{\epsilon \left( M-M^{2}u_{c}\left( 1-a^{2}\right) \right)
^{2}}{u_{c}\left( 1-3Mu_{c}+2M^{2}u_{c}^{2}\left( 1-a^{2}\right) \right) }.
\end{equation}%
where $u_{c}=1/r_{c}$ is the circular orbit of the particle. From Eq. (38)
we note that for a physical acceptable motion the constraint $%
1-3Mu_{c}+2M^{2}u_{c}^{2}\left( 1-a^{2}\right) >0$ must be satisfied.

\subsection{Null geodesics $\protect\epsilon =0.$}

Choosing $\epsilon =0$ in Eq. (36) and (37) we obtain 
\begin{equation}
\frac{E^{2}}{\ell ^{2}}=\frac{4\left( 1-a^{2}\right) +\sqrt{1+8a^{2}}-3}{%
128M^{2}\left( 1-a^{2}\right) }\left( 3-\sqrt{1+8a^{2}}\right) ^{2},
\end{equation}%
We can obtain the minimum radius for circular orbits $u_{cm}$ as%
\begin{equation}
u_{cm}=\frac{3-\sqrt{1+8a^{2}}}{4M\left( 1-a^{2}\right) }.
\end{equation}%
The corresponding value of $r_{cm}$ is given by\ 
\begin{equation}
r_{c}=\frac{3M+M\sqrt{1+8a^{2}}}{2}
\end{equation}%
We conclude from Eq.(42) that there is only one equilibrium circular orbit
for the photon with the ratio given by Eq.(40). The limiting case $a=1,$%
implies $r_{cm}=3M$ \ which corresponds to SBH limit. We observe that at
this radius $r_{cm}$ the geodesic equations allow an unstable circular
orbit. Plots of Eqs. (40) and (42) in terms of the external parameter $a$\
are shown in figures 4 and 5 respectively. The figures show that for larger
value of $a$ the circular orbit of the photon has larger radius but smaller
value of $\left( E^{2}/\ell ^{2}\right) /M^{2}.$ Note that the maximum value
of $r_{cm}$\ is $3M$ \ which implies that the circular orbit of the photon
in SEBH\ is smaller (closer to the singularity) compared with SBH as seen in
Fig. 5.

To obtain the effective potential in the case of null geodesic we use Eq.
(15) which can be rearrange as%
\begin{equation}
\left( \frac{dr}{d\widetilde{s}}\right) ^{2}+V_{eff.}=\varepsilon _{eff.},
\end{equation}%
where $\varepsilon _{eff.}=$ $E^{2}/\ell ^{2},$ we have replace $s=%
\widetilde{s}$ and 
\begin{equation}
V_{eff.}=\frac{1}{r^{2}}\left( 1-\frac{2M}{r}+\frac{M^{2}}{r^{2}}\left(
1-a^{2}\right) \right) .
\end{equation}%
Again for stable circular orbit the conditions are $V_{eff.}^{\prime }\left(
r=r_{c}\right) =0$ and $V_{eff.}^{\prime \prime }\left( r=r_{c}\right) >0.$
The effective potential is plotted in Fig. 6 which shows that the circular
trajectories for photons are unstable. Therefore no stable circular orbits
exist for photons.

On the other hand, the radial equation can be reduced to 
\begin{equation}
\left( \frac{du}{d\phi }\right) ^{2}=\frac{1}{D^{2}}-\left(
1-2Mu+M^{2}\left( 1-a^{2}\right) u^{2}\right) u^{2}=g\left( u\right) .
\end{equation}%
where $D=\ell /E$ is the impact parameter. It is usual to study the null
geodesics depending on the value of $D$ [34,15,16]. We note that, the
quartic equation $g\left( u\right) =0,$ has four roots, at least two real
roots. We shall be interested when the other two remaining roots are either
complex or real. Let $D=D_{c}=\frac{r^{2}}{r^{2}-2Mr+M^{2}\left(
1-a^{2}\right) },$ where $D_{c}$ denotes for which $g\left( u\right) =0$ has
double roots. Then, when $D>$\bigskip $D_{c}$, we have orbits of two kinds:
first kind orbits where photon coming from $\infty $ and receding back to $%
\infty $ and second kind orbits having two turning points. When $D<D_{c}$
then $g\left( u\right) =0,$ has only one real root, photon coming from $%
\infty $, crossing the horizon and have a turning point for a finite value
of $r<r_{h}$\bigskip . We must note that Eq. (45) is similar to the case of
RN geometry with the replacement of $\left( \text{the charge }%
Q^{2}=M^{2}\left( 1-a^{2}\right) \right) ,$ therefore when $D=D_{c}$ \ we
adapt the solution for $\phi $ [8]: 
\begin{equation}
\pm \phi =\frac{1}{\sqrt{A}}\log \left[ 2\left( A\sqrt{A\xi ^{2}+M\left(
B-1\right) \xi -M^{2}\left( 1-a^{2}\right) }\right) +2A\xi +M\left(
B-1\right) \right] ,
\end{equation}

where 
\begin{eqnarray}
A &=&\left( \frac{3-B}{4\left( 1-a^{2}\right) }\right) B>0,\text{ since }%
0<a\leq 1 \\
\qquad B &=&\sqrt{1+8a^{2}},\qquad \xi =\frac{1}{u-u_{cm}}.  \notag
\end{eqnarray}%
The massless particle circular orbit for the critical $D_{c}$ is given in
Fig. 7 for $a=0.7.$ It is seen that photons have unstable circular orbits
which is similar to the case of RNBH where the photons just barely cross the
horizon before they are terminated. \bigskip

\subsection{Timelike geodesics $\protect\epsilon =1.$}

Choosing $\epsilon =1.$ in Eq. (38) and (39) we obtain 
\begin{equation}
E^{2}=\frac{\left( 1-2Mu_{c}+M^{2}u_{c}^{2}\left( 1-a^{2}\right) \right) ^{2}%
}{1-3Mu_{c}+2M^{2}u_{c}^{2}\left( 1-a^{2}\right) },
\end{equation}%
\begin{equation}
\ell ^{2}=\frac{\left( M-M^{2}u_{c}\left( 1-a^{2}\right) \right) ^{2}}{%
u_{c}\left( 1-3Mu_{c}+2M^{2}u_{c}^{2}\left( 1-a^{2}\right) \right) }.
\end{equation}%
For a physical acceptable motion the condition $1-3Mu_{c}+2M^{2}u_{c}^{2}%
\left( 1-a^{2}\right) >0$ is required. Which is exactly the same radius of
the unstable circular orbit Eq. (41) as in the case of null geodesics. The
radial dependence of both $E^{2}$ and $\ell ^{2}$ of the test particle
moving on circular orbits are given in figures. 8 and 9 respectively. The
figures show that as the external parameter $a$ increases both the energy
and the angular momentum values increase until they reach the maximum values
which is when $a=1$ (the case of SBH). It is a consequence of the coupling
of S solution and BR solution that, both values of $E^{2}$ and $\ell ^{2}$
of the test particle moving on circular orbits\ having less values than the
SBH case.

To exam the stability of the equilibrium circular motion of a massive
particle we plot the effective potential (given in Eq. (18) with $\epsilon =1
$) in terms of $r/M$ for various values of $\ell ^{2}/M^{2}$. The behavior
of the effective potentials of SEBH and SBH are plotted in figures 10 and
11, respectively. We notice that both figures are almost similar. They are
concave down and become asymptotically constant. Therefore, even for given
larger energy than the asymptotic value of the effective potential particle
in SEBH would escape to infinity as in the case of SBH. The only difference
between the two figures is that the effective potential of the SEBH is
steeply rising more than SBH.

Using the values of $E^{2}$ and $\ell ^{2}$ given in Eqs. (48) and (49) the
radial equation in timelike geodesic can be reduced to 
\begin{eqnarray}
\left( \frac{du}{d\phi }\right) ^{2} &=&\left( u-u_{cm}\right) (\left[
M-M^{2}\left( 1-a^{2}\right) u_{cm}-\frac{M^{2}}{\ell ^{2}u_{cm}^{2}}\right]
u_{cm}  \notag \\
&&+2\left[ M-M^{2}\left( 1-a^{2}\right) u_{cm}\right] u-M^{2}\left(
1-a^{2}\right) u^{2})u^{2}.
\end{eqnarray}

We notice that the solution of the above equation (50) is similar to the
null geodesics Eq. (46) with the change of 
\begin{equation}
A=\left( \frac{3-\sqrt{1+8a^{2}}}{4\left( 1-a^{2}\right) }\right) \sqrt{%
1+8a^{2}}-\frac{M^{2}}{\ell ^{2}u_{cm}^{2}}.
\end{equation}%
The above solution provides us with second kind orbits. Figures 12 and 13,
respectively, represent the second kind orbits associated with a stable and
unstable circular orbits.

It is known that there exists a minimal radius at which stable circular
motion is still possible, this orbit is called ISCO. In order to find $%
r_{ISCO}$ we remind that to have a stable circular orbit we must supplement
the conditions $g\left( u\right) =g^{\prime }\left( u\right) =0,$ with the
equation $g^{\prime \prime }\left( u\right) =0.$ therefore one can easily
obtain 
\begin{equation}
4M^{3}\left( 1-a^{2}\right) ^{2}u^{3}-9M^{2}\left( 1-a^{2}\right)
u^{2}+6Mu-1=0,
\end{equation}%
which has the solution 
\begin{equation}
r_{ISCO}=\frac{4M\left( 1-a^{2}\right) }{3-\sqrt[3]{1+8a^{2}+4a\sqrt{1+4a^{2}%
}}-\sqrt[3]{1+8a^{2}-4a\sqrt{1+4a^{2}}}},
\end{equation}%
Expansion of Eq. (53) in degrees of $\left( 1-a^{2}\right) $ we obtain 
\begin{equation}
r_{ISCO}\cong 6M-\frac{3}{2}M\left( 1-a^{2}\right) +\frac{M}{127}\left(
1-a^{2}\right) ^{2}+order\text{ }of\text{ }\left( M\left( 1-a^{2}\right)
^{3}\right) ,
\end{equation}%
It is shown in the expansion above that the critical value of the $ISCO$
radii is $r_{ISCO}=6M,$ which corresponds to a test particle in the S
spacetime. We also note that $r_{ISCO}$\ is always smaller than $6M$ .\ To
justify our claim we have plotted Eqs. (53) in Fig. 14 showing the
dependence of the \ lower limit for the radius of $ISCO$ around a SEBH from
the external parameter $a$. It is shown in Fig. 14 that the radius of $ISCO$
is always smaller than $6M$ and as $a$ increases the radius of $ISCO$
increases. Therefore by decreasing $a$ from $1$ to a very close to $0$
(recall that $0<a\leq 1$) the radius is shifted very close to the horizon.

\bigskip

\section{Conclusion}

In this paper, we have described a detailed analysis of the geodesic motion
of both massless and massive particles in the spacetime of SEBH that
represent a static mass imbedded in a uniform em field. The metric of SEBH
combines both S and BR solutions in a single metric via the external
parameter $a$. As a result of this combination the horizon shrinks and the
geodesics are modified. We have investigated the effect of coupling S with
BR on the trajectories of the null and timelike geodesics. The paper is
separated in two parts, radial geodesics and circular geodesics. For the
radial geodesics, it is found that all radial orbits are stable since the
effective potential always has a minimum for any value of $a$. The effective
potential in the radial motion shows that, there is no upper bound for the
radial motion of the particle which means that particle can escape to
infinity. We have obtained exact analytical solutions for both proper time
and coordinate time. In radial null geodesics the obtained solutions are
similar to the S case. In radial timelike geodesics, it is shown that
particles falling toward the horizon in a finite proper time in SEBH take
longer than in the case of SBH.

The second part of our study was about circular geodesics. By analyzing the
effective potential we found all possible orbits allowed. It is shown that
there are no stable circular orbits for photons. However, for massive
particle we have obtained second kind orbits similar to the case of RNBH. We
have obtained the exact expression for $ISCO$ which corresponds to minimal
possible radius of stable circular orbits. Then we have plotted the
dependence of $r_{ISCO}$ and the $r_{cm}$\ on the external parameter $a$. We
have shown that both radii ( $r_{cm}$ and $ISCO)$ are shifted close to the
central object. We can conclude that the coupling of SBH with em field
affected all the trajectories of the null and timelike geodesics. Comparing
with SBH (without coupling with em field) we can summarize the effects as,
radial orbits are more stable, delay of particle falling towards the horizon
in the radial motion, no stable circular orbits for photons, stable and
unstable circular orbits for massive particle exist, and both the circular
orbits $r_{cm}$\ and the radius of $ISCO$ shrink and become close to the
central object. Finally, in future study we are going to extend our analysis
to study the geodesics equations for a spinning test particle since the
particle's orbit and motion will be different than the case of discussed in
this paper. 

\bigskip

\bigskip {\LARGE References}

[1] Robinson I 1959 Bull. Acad. Pol. Sci. \textbf{7},351.

[2] Bertotti B 1959 Phys. Rev. \textbf{116} 1331.

[3] Halilsoy M and Al-Badawi A 1998 IL Nuovo Cimento B \textbf{113} 761.

[4] Halilsoy M and Al-Badawi A1995 Class. Quantum Grav. \textbf{12} 3013.

[5] Halilsoy M 1993 Gen. Relativ. Gravit. \textbf{25} 975.

[6] George A Alekseev and Alberto A Garcia 1996 Phys. Rev. D \textbf{53},
1853.

[7] Halilsoy M 1993 Gen. Relativ. Gravit. \textbf{25} 275.

[8] Chandrasekhar S 1983 The Mathematical Theory of Black Holes Clarendon,
London.

[9] C. Chakraborty, 2014 Eur. Phys. J. C 74, 2759.

[10] P. Pradhan, 2015 Class. Quant. Grav. 32, 165001.

[11] G. V. Kraniotis and S. B. Whitehouse, 2003 Class. Quantum Grav. \textbf{%
20} 4817.

[12] G. V. Kraniotis, 2004 Class. Quantum Grav. \textbf{21} 4743.

[13] Cruz, N., Olivares, M., Villanueva, J.R. 2005 Class. Quantum Grav. 
\textbf{22} 1167.

[14] Sheng Zhou, Juhua Chen, Yongjiu Wang, 2012 Int. J. Mod. Phys. D, 21,
1250077.

[15] Stuchh'k, Z. and Hledik S 2002 Acta phys. Slov. 52 363.

[16] Stuchh'k, Z. and Slany P 2004, Phys. Rev. D 69 064001.

[17] Podolsky, 1999 J. Gen. Rel. Grav. 31, 1703.

[18] Eva Hackmann and Claus Lammerzahl 2008 Phys. Rev. D 78 024035.

[19] Halilsoy M , Gurtug O. and Mazharimousavi S. 2013 Gen. Relativ. Gravit. 
\textbf{45} 2363.

[20] Abdujabbarov A and Ahmedov B 2010, Phys. Rev. D 81 044022.

[21] S. A. Kaplan, 1949 JETP, 19, 951.

[22] L. D. Landau, E. M. Lifshitz, The Classical Theory of Fields (Pergamon,
Oxford, 1993).

[23] R. Ruffini and J. Wheeler, in Proceedings of the Conference on Space
Physics-Paris (ESRO, Paris, 1971).

[24] J. M. Bardeen, W. H. Press, and S. A. Teukolsky, 1972 Astrophys.J. 178,
347.

[25] P. I. Jefremov, O. Yu. Tsupko, G. S. Bisnovatyi-Kogan, 2015 Phys. Rev.
D. \textbf{91}, 124030.

[26] Hackmann, E., L\"{a}mmerzahl, C., 2008a Phys. Rev. Lett. \textbf{100},
171101.

[27] Hackmann, E., L\"{a}mmerzahl, C., 2008b Phys. Rev. D. \textbf{78},
024035.

[28] Abdujabbarov A, Ahmedov B and Hakimov A 2011, Phys. Rev. D 83 044053.

[29] Hackmann, E.,Kagramanova, V., Kunz, J., L\"{a}mmerzahl, 2008 C. Phys.
Rev. D. \textbf{78}, 124018.

[30] Dahia, F., Romero, C., da Silva, L.F.P., Tavakol, R. 2007 J. Math.
Phys. \textbf{48} 072501.

[31] Uzan, J. P., Lehoucq, R. 2001 Eur. J. Phys. \textbf{22} 317.

[32] Ali Ovgun, 2016 Int.J.Theor.Phys. 55 no.6, 2919-2927.

[33] Newman E Tand Penrose R 1962 J. Math. Phys. \textbf{3}
566.[Erratum-ibid. \textbf{4}, 998, 1963.

[34] Stuchh'k, Z. and Hledik S 1999, Phys. Rev. D 60 0044006.

\end{document}